\documentclass[lettersize,journal]{IEEEtran}
\usepackage[T1]{fontenc}
\usepackage{amsmath,amsfonts}
\usepackage{algorithmic}
\usepackage{algorithm}
\usepackage{array}
\usepackage{multirow} 

\usepackage{booktabs}
\usepackage{subcaption}
\usepackage{textcomp}
\usepackage{stfloats}
\usepackage{url}
\usepackage{verbatim}
\usepackage{graphicx}
\usepackage{bm}
\usepackage{pifont}
\usepackage{color}
\usepackage[colorlinks=true, urlcolor=blue, citecolor=blue]{hyperref}
\usepackage[dvipsnames]{xcolor}
\usepackage{algorithm}
\usepackage{algorithmic}
\usepackage{caption}
\usepackage{changepage}
\usepackage{tabularx}
\usepackage{array}
\usepackage[square,sort,comma,numbers]{natbib}
\usepackage{braket}

\newcommand{\revision}[1]{\textcolor{black}{#1}}
\begin{document}

\title{QAISim: A Toolkit for Modeling and Simulation of AI in Quantum Cloud Computing Environments}

\author{
\IEEEauthorblockN{Irwindeep Singh\IEEEauthorrefmark{1}\IEEEauthorrefmark{2}, Sukhpal Singh Gill\IEEEauthorrefmark{2}\thanks{Corresponding author: Sukhpal Singh Gill, School of Electronic Engineering and Computer Science, Queen Mary University of London, London, E1 4NS, UK. Email: s.s.gill@qmul.ac.uk}, Jinzhao Sun\IEEEauthorrefmark{3}, Jan Mol\IEEEauthorrefmark{3}} \\
\IEEEauthorblockA{\IEEEauthorrefmark{1}Department of Computer Science and Engineering, Indian Institute of Technology, Jodhpur, Rajasthan, India} \\
\IEEEauthorblockA{\IEEEauthorrefmark{2}School of Electronic Engineering and Computer Science, Queen Mary University of London, London, UK } \\
\IEEEauthorblockA{\IEEEauthorrefmark{3}School of Physical and Chemical Sciences, Queen Mary University of London, London, UK \\
Emails: b22ai022@iitj.ac.in; s.s.gill@qmul.ac.uk; jinzhao.sun@qmul.ac.uk; j.mol@qmul.ac.uk}
}

\markboth{Cluster Computing Springer, Jan~2026}%
{Shell \MakeLowercase{\textit{et al.}}: Bare Demo of IEEEtran.cls for IEEE Journals}

\maketitle

\begin{abstract}
Quantum computing offers new ways to explore the theory of computation via the laws of quantum mechanics. Due to the rising demand for quantum computing resources, there is growing interest in developing cloud-based quantum resource sharing platforms that enable researchers to test and execute their algorithms on real quantum hardware. These cloud-based systems face a fundamental challenge in efficiently allocating quantum hardware resources to fulfill the growing computational demand of modern Internet of Things (IoT) applications. So far, attempts have been made in order to make efficient resource allocation, ranging from heuristic-based solutions to machine learning. In this work, we employ quantum reinforcement learning based on parameterized quantum circuits to address the resource allocation problem to support large IoT networks. We propose a python-based toolkit called QAISim for the simulation and modeling of Quantum Artificial Intelligence (QAI) models for designing resource management policies in quantum cloud environments. We have simulated policy gradient and Deep Q-Learning algorithms for reinforcement learning. QAISim exhibits a substantial reduction in model complexity compared to its classical counterparts with fewer trainable variables.

\end{abstract}


\IEEEpeerreviewmaketitle

\section{Introduction}\label{sec:Inroduction}

\IEEEPARstart{Q}{uantum} computing is an emerging field with promising results for a wide range of modern Internet of Things (IoT) applications in the study of computer science and the theory of computation \cite{Transforming2024}.  
Quantum computing leverages the principles of superposition and entanglement and provides powerful tools for efficiently solving problems in many-body physics~\cite{Daley2022,yuan2021quantum,sun2023probing}, materials science~\cite{babbush2018lowdepth,cao2023ab}, and quantum chemistry~\cite{lee2021even}. This gives rise to the near-term quantum computation era, where we have noisy intermediate-scale quantum (NISQ) devices that perform quantum operations with some error \cite{Bharti2022,Chen2023}. However, physical quantum computers as we know today usually require very special environmental conditions to operate \cite{nguyen2024quantumcloudcomputingreview}. Due to this reason, NISQ devices are hosted and managed on cloud platforms like IBM Quantum \cite{IBMQuantum}, AWS \cite{AWSQuantum}, and Microsoft Azure \cite{AzureQuantum}.
This is referred to as Quantum Computing as a Service (QCaaS or QaaS), a cloud-based platform that delivers quantum computing capabilities to end users. Since quantum computing predominantly operates using the quantum-circuit model, these services allow users to input quantum circuits for execution and receive the desired measurements as output. 

\begin{figure}[ht]
    \centering
    \includegraphics[width=\linewidth]{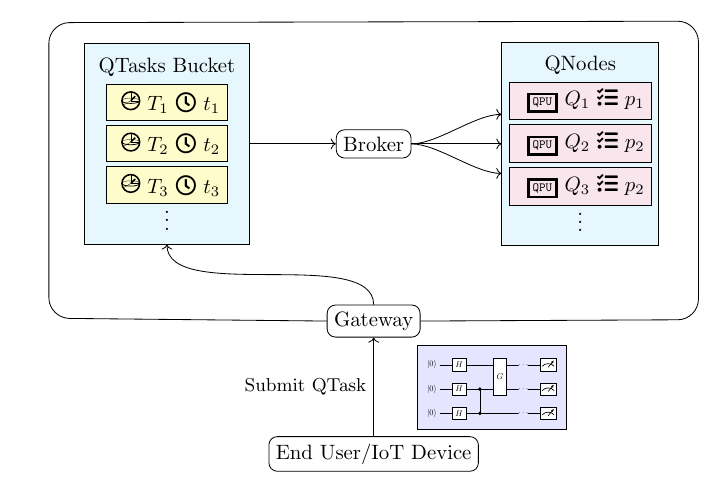}
    \caption{Top-Level abstraction of a quantum cloud computing environment and a sample task execution workflow.}
    \label{fig:qcc}
\end{figure}

Figure \ref{fig:qcc} gives a top-level abstraction of the working of a quantum cloud computing environment. \revision{The End User/IoT represents resource constrained applications that offload quantum tasks to the cloud to access quantum computational tasks. A gateway serves as an entry point to the platform, where submitted QTasks are placed into a QTask bucket. These tasks are then dispatched to available QNodes for execution by a broker.}

\subsection{Challenges}

Similar to any other cloud resource management system, quantum cloud environments also deal with the problem of efficient resource allocation with rising complexity of quantum processing units (QPUs) \cite{gill2025quantum}. Majority of solutions today uses carefully designed \textit{``static''} heuristics to address this problem. But the problem behind these heuristics is simple; first, these are designed for a specific problem i.e., a resource allocation algorithm for one system cannot be adopted on the other and second, these are static in nature i.e., with a change in the underlying cloud model, these heuristics become inefficient.

This trend of increasing hardware (or backend) complexity could also be seen in quantum computing systems. Quantum computers are becoming increasingly complex day by day with more efficient error correction strategies \cite{Campbell2024}. Designing heuristic-based algorithms to correctly manage these complex systems is next to impossible \cite{gill2022quantum}.

Furthermore, in a practical scenario, such a resource allocator needs to make online decisions under diverse conditions to fulfil the growing computational demand of modern IoT applications \cite{golec2024quantum}. These systems are hard to be adapted by algorithms. This demands for efficient \textit{``dynamic''} solutions that can be adapted to different systems and scenarios and online decision making. One possible solution is to utilize classical machine learning/reinforcement learning models to perform resource management \cite{Zhou2024}. However, these models can become very large with increasing number of parameters, making inference hard for researchers.

\subsection{Existing Solutions}

iQuantum \cite{iQuantum} is a discrete-event simulation toolkit designed to model quantum computing resources within cloud environments. Built upon the CloudSim framework \cite{CloudSim}, iQuantum enables researchers to design and evaluate quantum resource management policies. While this solves some issues in the existing frameworks \cite{QuNetSim, Coopmans2021}, it has its own limitations of heuristics-based resource management policies. Also, iQuantum was proposed to be a Java based solution, which might pose scalability issues considering most of the quantum computing community is python based.

To solve these problems, a new python-based toolkit QSimPy \cite{qsimpy} was proposed. It integrates with the Gymnasium environment to support the development and evaluation of reinforcement learning (RL) techniques aimed at optimizing quantum cloud resource management. While this solves all issues addressed earlier, however classical reinforcement learning frameworks are complex and large for inference purposes.

\subsection{Motivation and Our Contributions}

This paper explores the vast domain of quantum cloud (or any general cloud environment) resource management using quantum reinforcement learning strategies. Furthermore, we introduce \textbf{QAISim}, a python-based framework that offers to provide efficient resource management policies in quantum cloud environments. QAISim explores the advantages quantum computing has to offer in resource management. 
The \textbf{\textit{main contributions}} of this work are:
\begin{itemize}
    \item We proposed a new toolkit called QAISim that addresses the challenge of resource management in a dynamic quantum cloud computing environment. Leveraging advanced algorithms and simulations, QAISim enables seamless workload balancing for quantum hardware interfaces with IoT infrastructures.
    \item We integrated advanced RL algorithms such as Deep Q-Learning and Policy Gradient using parametrized quantum circuits (PQC) to efficiently manage resources within the QAISim environment, utilizing Quantum Machine Learning (QML) for intelligent decision making.
    \item We used Cirq for simulating quantum circuits and \revision{their} properties and performed Quantum Reinforcement Learning (QRL) tasks using extensive QML support provided by TensorFlow Quantum.
    QAISim also integrates python libraries like \revision{gymnasium \cite{gymnasium}} to create RL environment for workload simulation.
    \item We compared the performance of QAISim with \revision{baselines,} a heuristics-based greedy algorithm and a classical deep reinforcement learning algorithms and evaluated the performance in terms of mean difference of the rewards over a given set of episodes. Experimental results indicate that QAISim is able to achieve state-of-the-art results with QRL settings.
\end{itemize}

\subsection{Article Organization}
The rest of the paper is organized as follows: Section \ref{sec:Related} presents related work. Section \ref{sec:Background} discusses background, including definitions and concepts. Section \ref{sec:Architecture} presents the architecture of QAISim. Section \ref{sec:Implementation} discusses the design and implementation of QAISim. Section \ref{sec:Experiments} presents the experimental setup and results. Finally, Section \ref{sec:Conclusion} concludes the paper and highlights future directions.

\section{Related Works}\label{sec:Related}

In this section, we introduce simulation tools dedicated to quantum cloud computing and subsequently conduct a comparative analysis, focusing on the distinct features and contributions of QAISim compared to these existing solutions. By evaluating the capabilities of QAISim, we aim to highlight its potential advantages and novel contributions to the field of quantum cloud simulation.

Nguyen et al. \cite{iQuantum} proposed iQuantum, the first of its kind to develop a quantum cloud simulation framework. iQuantum is a Java-based framework built upon CloudSim \cite{CloudSim} simulation architecture for simulation and modeling of hybrid quantum computing systems within the cloud-edge environment. The toolkit helps researchers with the development and evaluation of resource management strategies such as task scheduling, backend selection, hybrid task orchestration, and offloading. However, this work cannot be leveraged if the underlying environment is changed.

To manage the problem of heuristic management in iQuantum, Nguyen et al. \cite{qsimpy} proposed QSimPy, a learning-centric framework for quantum cloud resource management. QSimPy utilized deep reinforcement learning to train a deep-q network (DQN) for designing efficient resource management policies. This work helps researchers train similar networks based on the underlying environment. However, these networks often grow very large based on the depth of the network and hence can be difficult to scale.

Apart from these works, a variety of simulation frameworks are known with a focus towards specific applications. For example, DRAS-CQSim \cite{dras_cqsim} employs reinforcement learning for efficient scheduling in High-Performance Computing (HPC) clusters. However, this work does not provide direct quantum resource management applications. Other than this, qgym proposed by Linde et al. \cite{qgym} provides a benchmarking for RL-based quantum circuit compilation, which helps in efficiently mapping circuits to a given quantum hardware. However, no resource management was done in this as well.

\subsection{Critical analysis}

Table \ref{table:comparison_table} shows the comparison of QAISim with existing frameworks and simulators. Our research paper stands out in the field due to its innovative approach, which specifically targets resource management quantum computing systems using QRL models. To harness the full potential of quantum computing, it necessitates advanced simulation tools capable of addressing its unique challenges to simulate QRL based resource management \revision{policies} for quantum computing systems. Furthermore, our proposed simulator, QAISim has ability to simulate the full spectrum of quantum computing functionalities and assess different aspects of such a system positions it as a promising choice for researchers seeking to conduct in-depth assessments of quantum computing systems and modern IoT applications, making it stand out in the landscape of quantum computing simulation tools.

\begin{table*}
\caption{\small \textcolor{black}{Comparison of QAISim with existing works. $\times$:= method does not support the property}}
\label{table:comparison_table}
\begin{center}
\footnotesize
\begin{tabular}{|c||c|c|c|c|c|c|}
    \hline
    {\textbf{Work}} & \textbf{AI} & \textbf{Resource Management}  & \textbf{QRL Algorithms} & \textbf{Quantum Cloud} & \textbf{QPU Metrics} & \textbf{IoT}  \\
    \hline \hline
    DRAS-CQSim \cite{dras_cqsim} & \checkmark & \checkmark & X & X  &  X &  X \\ \hline
    qgym \cite{qgym} & \checkmark & X & X & X & X &  X\\  \hline
    iQuantum \cite{iQuantum} & X & \checkmark & X &	\checkmark & \checkmark &  X \\ \hline
    QSimPy \cite{qsimpy} &  \checkmark & \checkmark & X  &  \checkmark & \checkmark &  X \\ \hline
    \textbf{QAISim (this work)} & \checkmark & \checkmark & \checkmark & \checkmark & \checkmark & \checkmark \\ \hline
\end{tabular}
\end{center}
\end{table*}

\section{\textcolor{black}{Background: Definitions and Concepts}}\label{sec:Background}
This section discusses important definitions and concepts for understanding this work. \revision{Table \ref{tab:symbols} provides the symbols and definitions used in this research work.}

\begin{table}[ht!]
    \centering
    \caption{\revision{Symbols and definitions}}
    \label{tab:symbols}
    \begin{tabular}{@{}>{\color{black}}l>{\color{black}}l@{}}
        \toprule
        \textbf{Symbol} & \textbf{Definition} \\
        \midrule
        $\mathbf{H}$ & Complex vector space used in hilbert space \\ 
        H & Hadamard Gate \\
        $R_{x}(\phi)$ & Single qubit rotation gate through angle $\phi$ across x-axis \\
        $R_{y}(\phi)$ & Single qubit rotation gate through angle $\phi$ across y-axis \\
        $R_{z}(\phi)$ & Single qubit rotation gate through angle $\phi$ across z-axis \\
        CZ & Controlled Z Gate \\
        CNOT & Controlled NOT Gate \\
        SWAP & SWAP Gate \\
        PQC & Parametrized Quantum Circuit \\
        $M_{m}$ & Measurement operator for outcome $m$ \\
        $O$ & Quantum observable for measurement \\
        $\braket{O}$ & Expected value of the observable $O$ w.r.t. $\ket{\psi}$ \\
        $U_{\text{var}}(\phi)$ & Variational layer in PQC with parameters $\phi$ \\
        $U_{\text{enc}}(s, \lambda)$ & Encoding layer in PQC with state $s$ and parameters $\lambda$ \\
        \bottomrule
    \end{tabular}
\end{table}

\subsection{Quantum Computing and Parametrized Quantum Circuits}
This section gives a brief introduction to the basics of quantum computing and parametrized quantum circuits. We closely follow definitions from the textbook by Nielsen and Chuang \cite{Nielsen_Chuang_2010}. We use the Dirac notation to represent a state vector as $\ket{\psi}$ and its conjugate transpose as $\bra{\psi}$.

In quantum systems, the fundamental unit of information is the qubit, analogous to the bit in classical systems. Formally, a qubit is represented as a unit vector within a two-dimensional Hilbert space. Here, a Hilbert space is defined as a pair $(\mathbf{H}, \braket{|})$, where $\mathbf{H}$ is a complex vector space and $\braket{|}:\mathbf{H} \times \mathbf{H} \to \mathbb{C}$ is a map called inner product on $\mathbf{H}$. Similar to any vector space, a qubit can be represented as a linear combination of its basis states. The basis states for a qubit are denoted as $\ket{0}$ and $\ket{1}$, which are defined as follows:

\begin{equation}
    \ket{0} = \begin{pmatrix} 1\\ 0\end{pmatrix} \quad \text{and} \quad \ket{1} = \begin{pmatrix} 0\\ 1\end{pmatrix}
\end{equation}

Based on this, we define a quantum state to be a superposition of the basis states as follows:

\begin{equation}
    \ket{\psi} = \alpha \ket{0} + \beta \ket{1}, \qquad \alpha, \beta \in \mathbb{C}
\end{equation}
\begin{equation}
    |\alpha|^{2} + |\beta|^{2} = 1
\end{equation}

\begin{figure*}
    \centering
    \begin{subfigure}[c]{0.5\textwidth}
        \includegraphics[width=\linewidth]{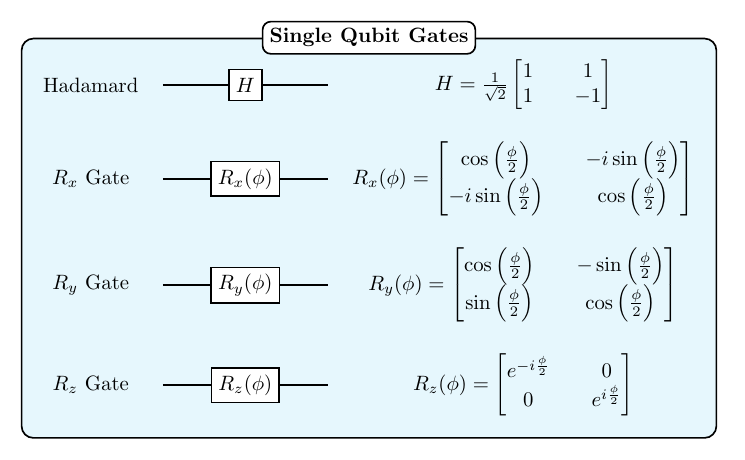}
        \caption{Examples of single qubit gates.}
        \label{fig:one_qubit}
    \end{subfigure} \hfill
    \begin{subfigure}[c]{0.45\textwidth}
        \includegraphics[width=\linewidth]{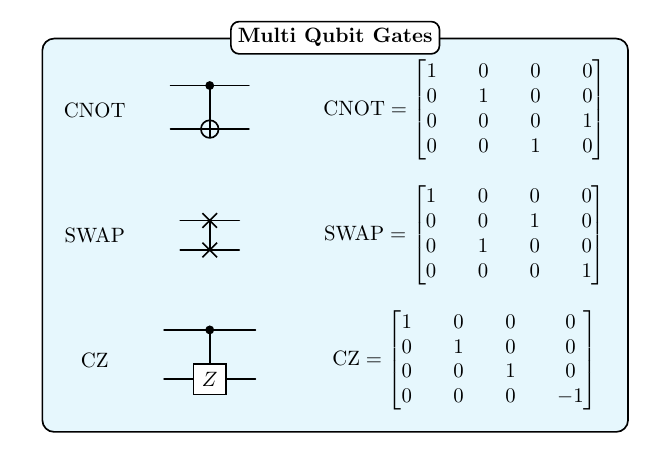}
        \caption{Examples of two qubit gates.}
        \label{fig:two_qubit}
    \end{subfigure}
    \caption{Important quantum gates (unitary operators) including hadamard, control and rotation gates.
    }
    \label{fig:quantum_gates}
\end{figure*}

In a quantum mechanical system, the evolution of a state is governed by the time-dependent Schr$\ddot{\text{o}}$dinger equation. In its discrete representation, this evolution is expressed as a Unitary Transformation, $\ket{\psi_{t_1}} = U \ket{\psi_{t_{0}}}$ where $U$ is a unitary operator (or gate). We represent these quantum operations using a tensor network (or a quantum circuit). Figure \ref{fig:quantum_gates} represents some important one and two-qubit quantum gates on a quantum circuit. For a system of qubits, we compute tensor product of all the states to arrive at the final state.

Quantum measurements, in contrast, are characterized by a set $\{M_{m}\}$ of measurement operators corresponding to each of $m$ possible outcomes of the measurement (example, either $\ket{0}$ or $\ket{1}$ for every qubit in the system, hence $2^{n}$ possibilities for $n$-qubit system). Once a qubit is measured, further measurements would always result in the same outcome.

Building upon quantum measurements, we now introduce observables. In quantum mechanics, observables are defined by Hermitian operators acting on the qubits. For an observable $O$, its spectral decomposition $O = \sum_{m} \lambda_{m} \ket{\phi_{m}}\bra{\phi_{m}}$ in terms of its eigenvalues $\lambda_{m}$ and eigenstates $\ket{\phi_{m}}$ defines the outcome of a measurement i.e., 
$M_{m} = \sqrt{p(m)}\ket{\phi_{m}}\bra{\phi_{m}}$  where $p(m) = |\braket{\phi_{m}|\psi}|^{2}$ is the probability of obtaining the outcome $m$. Expectation value of the observable with respect to $\ket{\psi}$ is given by $\braket{O} = \braket{\psi|O|\psi}$.

We will focus on Parametrized Quantum Circuits (PQCs) in this work. PQCs or Variational Quantum Circuits (VQCs) are quantum models with a set of trainable variables used in its operations  \cite{peruzzo2014variational,Ying2017,endo2020variational}. The quantum circuit is said to be parametrized by the trainable variables. Examples of such circuits are the rotation gates $R_{x}(\phi)$, $R_{y}(\phi)$ and $R_{z}(\phi)$ shown in Figure \ref{fig:quantum_gates}, parametrized by the rotation angle $\phi$. PQCs are widely used in most of the quantum machine learning algorithms~\cite{cerezo2021variational} and in experiments~\cite{guo2024experimental,robledo2025chemistry}, especially in quantum neural networks \cite{qnn}, and have shown promising results in theory \cite{Abbas2021}. In QNNs, we parametrize a quantum circuit with a set of parameters $\theta$ and based on some observations $\braket{O_{i}}$, we optimize the parameters $\theta$ using a classical optimizer.

\begin{figure*}
    \centering
    \includegraphics[width=\linewidth]{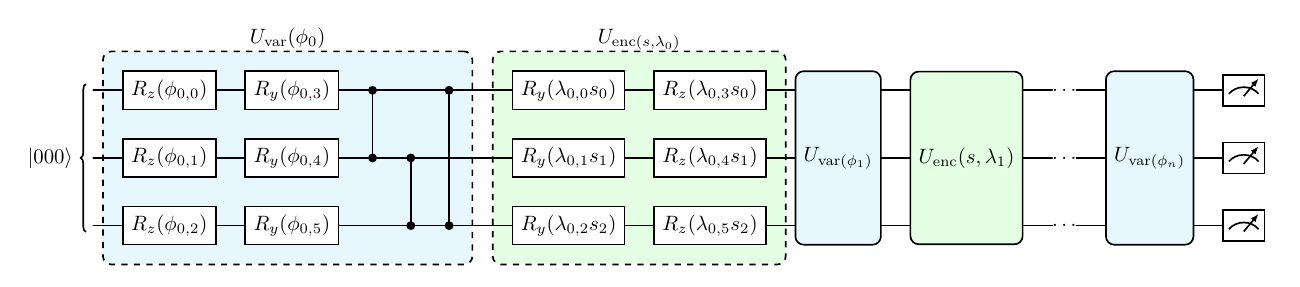}
    \caption{A 3-qubit data re-uploading parametrized quantum circuit with $U_{\text{var}}(\phi_{i})$ and $U_{\text{enc}}(s, \lambda_{i})$ layers with variational parameters $\phi_{i}$ (rotational angles) and $\lambda_{i}$ (scaling parameters) and input parameters $s$. We use CZ operators to perform the entangling operations in $U_{\text{var}}(\phi_{i})$.}
    \label{fig:pqc_rl}
\end{figure*}

In this work, we adopt the framework of using parametrised quantum circuits for reinforcement learning introduced by Jerbi et al. \cite{jerbi2021} (see Figure \ref{fig:pqc_rl}). We use this particular circuit for both policy gradient and Deep Q-Learning using data re-uploading \cite{reuploading}. 
As shown in Figure \ref{fig:pqc_rl}, the data re-uploading PQC consists of two different types of layers; $U_{\text{var}}(\phi_{i})$ layer performing rotation operations using angles $\phi_{i}$ independent of input parameters and a $U_{\text{enc}}(s, \lambda_{i})$ layer encoding input parameters using rotation operations with scaling parameters $\lambda_{i}$. Finally we compute the expectation values for each observable and optimize the trainable parameters accordingly. Section \ref{sec:Architecture} discusses the choice of observables and learning algorithms used based on this parametrized quantum circuit.

\begin{figure*}
    \centering
    \includegraphics[width=0.8\linewidth]{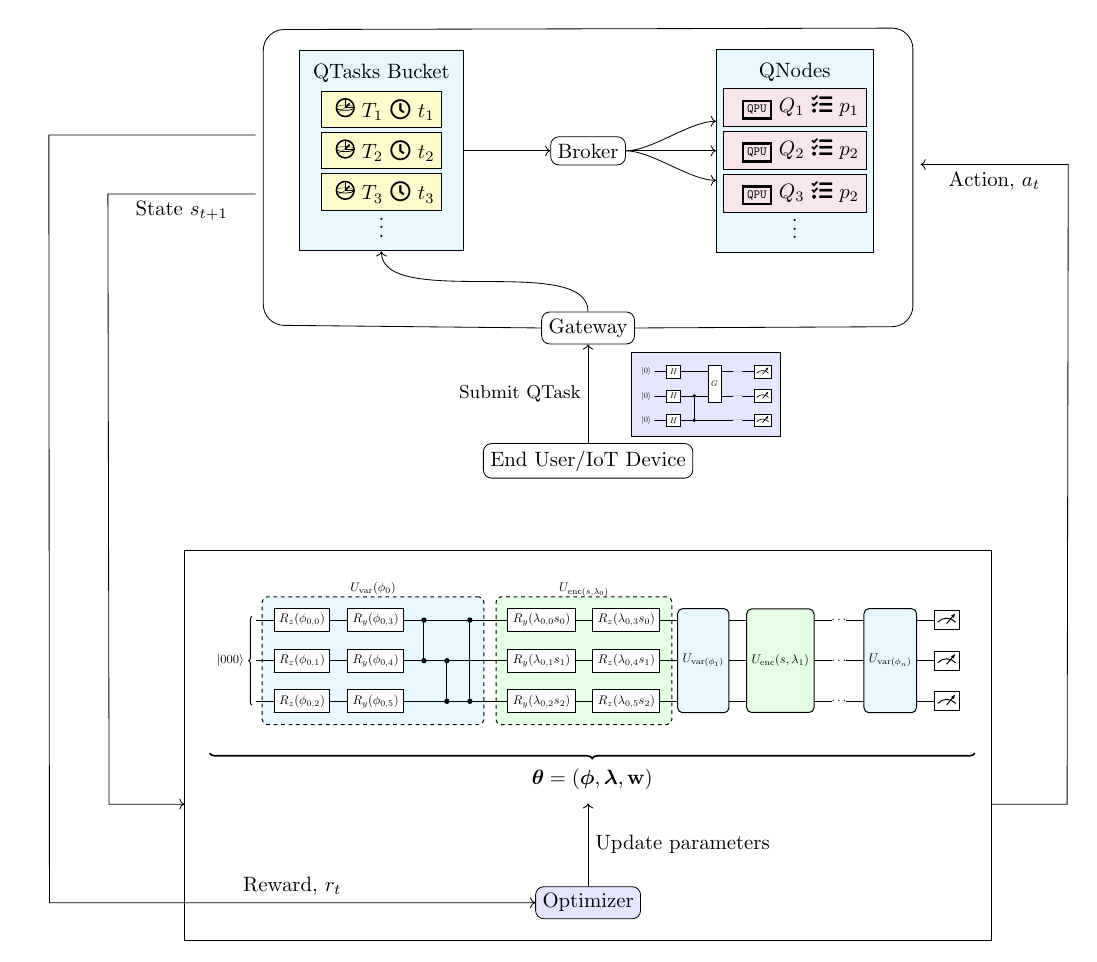}
    \caption{QAISim Architecture.}
    \label{fig:qaisim_arch}
\end{figure*}

\subsection{Resource Management in QC}\label{sec:resource_management}

Now, we formulate the reinforcement learning based resource management in QC. We describe a quantum computing system using QTasks (quantum tasks that need to be executed on hardware) and a set of QNodes (quantum processing units which process all the quantum tasks). Based on this, we define the state space, action space and rewards for resource management using reinforcement learning as follows:

\begin{enumerate}
    \item \textbf{State Space:} We define the state of a system as a feature vector containing the number of pending jobs in each QNode and the QTask metrics (like arrival time, number of qubits and circuit depth). Each different state of the RL (or QRL in this case) configuration would represent a QTask waiting to be mapped to some QNode.
    \item \textbf{Action Space:} Action space in any RL configuration is the set of all possible actions that can be taken from a given state $s$. We define the action space as the set of all QNodes as for each state, we can select a QNode and map the QTask to that QNode. Hence, our action space is a discrete set $\mathcal{A} = \{0, 1, 2, \dots, |Q|-1\}$, where $|Q|$ refers to the number of QNodes in the environment.
    \item \textbf{Rewards:} We define the rewards for each action (decision made by the QRL agent) to guide it towards good solutions for our objective (i.e., to minimize the average task completion time). For any action taken (QTask mapped to QNode), let $T$ be the total time from the arrival of QTask to its completion. The reward is given by:
    \begin{equation}
        R = \begin{cases} \frac{1}{T} , & \text{if QTask can be executed}\\ -10, & \text{otherwise} \end{cases}
    \end{equation}
    \revision{This function gives a simple, bounded reward that encourages faster task completion. Furthermore, Table \ref{tab:qpu_metrics} reports the mean reward obtained for task execution on a certain device using 1024 shots. A heavy penalty of -10 is applied for execution failures, which is significantly larger in magnitude compared to the positive rewards.}
\end{enumerate}

In addition, a survey paper on quantum reinforcement learning can be found by Meyer et al. \cite{qrl_survey} for a detailed explanation and rigorous proofs.

\section{QAISim  Architecture}\label{sec:Architecture}

Figure \ref{fig:qaisim_arch} shows the system architecture of QAISim. In QAISim, we extend the basic components of the QSimPy framework \cite{qsimpy} and develop new QAI-based simulation models for managing quantum resources. \revision{QAISim simulates a reinforcement learning environment for the quantum cloud environment and uses the PQC backbone to represent RL actions. Furthermore, the PQC contains trainable parameters that can be updated using an optimizer during learning. The End User/IoT represents resource constrained applications that offload quantum tasks to the cloud to access quantum computational tasks. A gateway serves as an entry point to the platform, where submitted QTasks are placed into a QTask bucket. These tasks are then dispatched to available QNodes for execution by a broker.} The basic components of the QAISim architecture are described in the following subsections.

\subsection{Quantum Processing Unit (QPU)}
Quantum Processing Units (QPUs) provide the quantum hardware required by any quantum task. We refer to these QPUs as QNodes in this work. Each QPU is associated with certain metrics such as CLOPS (Circuit Layer Operations Per Second), EPLG (Error per Layered Gate) \cite{eplg_paper} and Quantum Volume \cite{qvol_paper}. For simplicity, we assume that the execution time of a \revision{QTask} only depends on layers in \revision{QTask}, $L(\text{task})$ and CLOPS of the QNode. Hence, for a single shot run of any \revision{QTask}, its execution time is given as follows.

\begin{equation}
    \text{Execution time, } T = \frac{L(\text{task})}{\text{CLOPS(QPU)}}
\end{equation}

\subsection{Modeling of Tasks}
Quantum tasks are modeled as quantum circuits (defined using OpenQASM \cite{qasm} as inputs which are then transpiled to required QPUs using a qiskit-based \cite{qiskit} transpiler). Each QTask is associated with certain metrics such as circuit layers (representing number of layers in the circuit), gate counts (number of gates in the circuit) and shots (i.e., number of times the circuit has to be executed). Based on these metrics, each task has its definite time and resource requirements, which they will consume once any QNode is allotted to them.

\subsection{Modeling of QAI Models}
The QAI module consists of QRL-based resource management algorithms for effective handling of hardware resources, which is explained in section \ref{sec:Implementation}.

\section{Design and Implementation of QAISim}\label{sec:Implementation}

\subsection{\textcolor{black}{Design Description}\label{sec:Design}}
\begin{figure*}
    \centering
    \includegraphics[width=\linewidth]{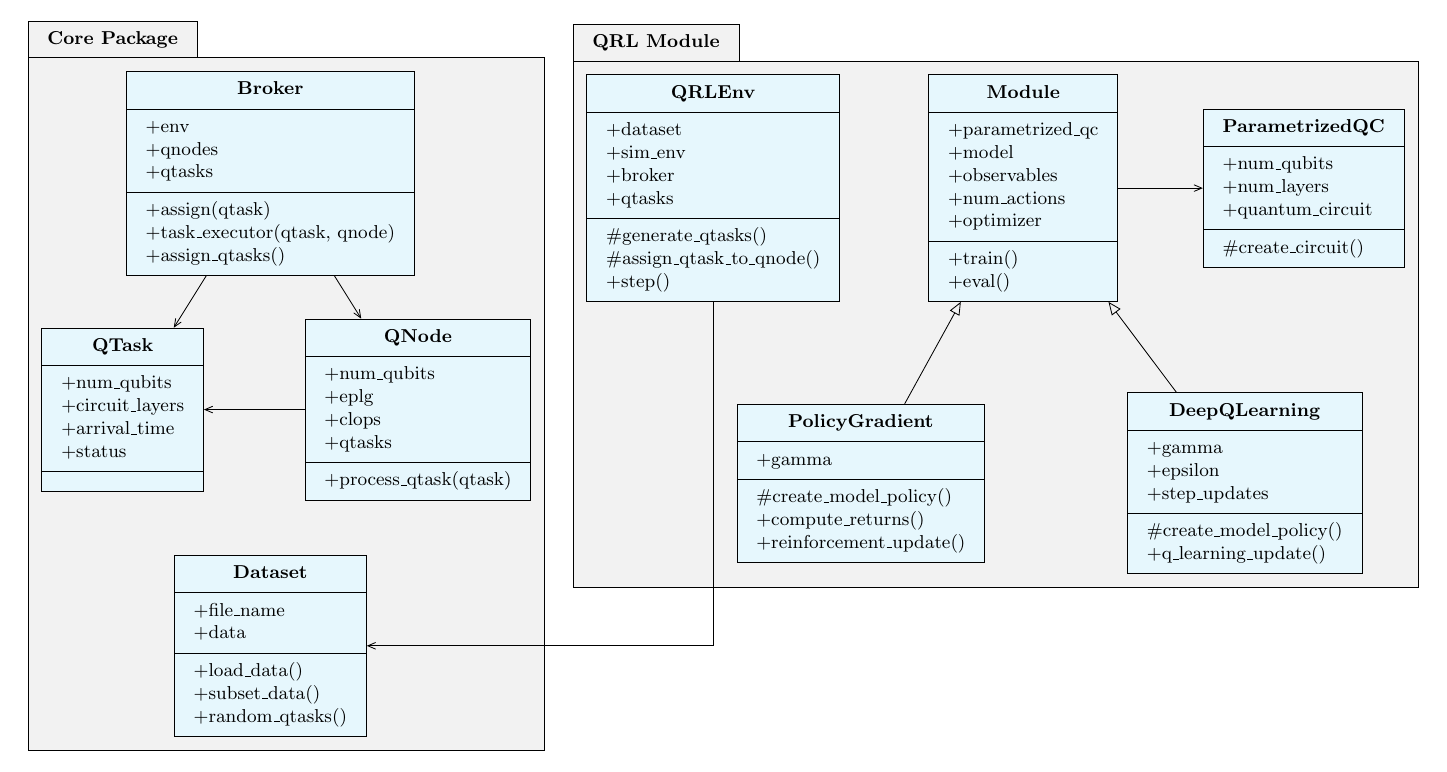}
    \caption{Fundamental Classes of QAISim.}
    \label{fig:class_dia}
\end{figure*}

The fundamental classes of QAISim are shown in Figure \ref{fig:class_dia}. The following are the main classes of QAISim.

\subsubsection{Core Package}
This class serves as the central orchestrator within the QAISim system. Its primary role is to facilitate the seamless coordination and operation of various integral components, including QNodes, QTasks, Broker, and more. Upon initialization, it instantiates and registers all available QTasks and QNodes present. Once these QNodes and QTasks are instantiated, the broker maintains a queue of pending tasks and a pool of free nodes. A QTask encapsulated a quantum workload and a QNode models a quantum processing unit (or QPU). Both of these are discussed in detail in section \ref{sec:resource_management}. Lastly, the broker aggregates QTasks and QNodes and find a mapping for each QTask to a QNode, our goal in this study is to train a quantum neural network, i.e., a parametrized quantum circuit to find such a mapping while ensuring efficiency in terms of training time.

\subsubsection{Quantum Circuits and QRL Integration}
This class of QAISim deals with the quantum logic and abstracts the task to node mapping function for the broker. The primary role of this class is to initialize, train and inference parametrized quantum circuits to be used for the decision making by the broker. The QAI module, implementing QRL algorithms for training are extended from this class. Section \ref{sec:qrl} discusses the modeling and implementation of QAI models for resource management.

\subsection{Modeling and Implementation of QAI models}\label{sec:qrl}
QAISim implements two reinforcement learning algorithms i.e., Policy Gradient and Deep Q-Learning to train parametrized quantum circuits. We use a baseline resource management model using a Greedy Algorithm, and furthermore, we compare the performance of our model with classical DRL agents proposed in QSimPy \cite{qsimpy}. As we have already discussed the RL formulation of the resource management problem, our RL agents aim to maximize the reward at each timestep.

\subsubsection{Baseline Model - Greedy Algorithm}
We have considered a greedy algorithm \cite{baseline} to develop a baseline model for QAISim. An algorithmic representation of the baseline is shown in Algorithm \ref{alg:greedy}. This is a simple resource management algorithm where we assign a given QTask to a QNode with the minimum number of pending tasks. This method allows us to focus on task total time by reducing the waiting time.

\begin{algorithm}[ht]
  \caption{Greedy Algorithm for resource management in Quantum Cloud}\label{alg:greedy}
  \begin{algorithmic}[1]
    \REQUIRE Input tasks $\mathcal{T}$, Set of QNodes $\mathcal{Q}$
    \STATE mapping $\leftarrow \{\}$
    \FOR{task $t \in \mathcal{T}$}
        \STATE valid\_qnodes $\leftarrow \{q \in \mathcal{Q} | \quad q.\text{n\_qubits} \ge t.\text{n\_qubits}\}$
        \STATE q\_mapped $\leftarrow$ QNode with least pending tasks among valid\_qnodes
        \STATE mapping[t] $\leftarrow$ q\_mapped
    \ENDFOR
    \RETURN mapping
  \end{algorithmic}
\end{algorithm}

\subsubsection{Policy Gradient}
Policy Gradient algorithm is one of the most fundamental reinforcement learning algorithm. In standard RL literature \cite{rl_book}, a policy, $\pi_{\mathcal{\theta}}(a|s)$ is defined as a strategy function for the agent, i.e., the agent works according to the policy outputs. In policy gradient methods, we try to learn this policy by performing gradient descent or similar gradient-based optimization algorithms on policy parameters. We utilize the softmax-PQC policy defined by Jerbi et al. \cite{jerbi2021}, in which the policy is given by a probability distribution derived from the expectation value of a PQC.

\begin{equation}
    \pi_{\boldsymbol{\theta}}(a|s) = \frac{e^{\braket{O_{a}}_{s, \boldsymbol{\theta}}}}{\sum_{a'} e^{\braket{O_{a'}}_{s, \boldsymbol{\theta}}}}
\end{equation}

Where, $\braket{O_{a}} = \braket{\psi_{s, \boldsymbol{\theta}} | w_{a} Z^{\otimes n} | \psi_{s, \boldsymbol{\theta}}}$ is the expectation value of the weighted $Z^{\otimes n}$ operator associated to the action $a \in \mathcal{A}$, and $Z$ denotes the Pauli-$Z$ gate. The weight $w_{a}$ is a trainable parameter, and the full set of model trainable parameters is $\boldsymbol{\theta} = (\boldsymbol{\phi}, \boldsymbol{\lambda}, \mathbf{w})$, where $\boldsymbol{\phi}$ and $\boldsymbol{\lambda}$ parametrize with the PQC and $\mathbf{w}$ parametrizes the policy weights.

We utilize the REINFORCE variant of the policy gradient method. Algorithm \ref{alg:policy_grad} represents the outline of the training algorithm used. This method calculates the expected returns $G_{t}$ at each timestep along the episode using a discount factor $\gamma \in (0, 1]$. These returns are then combined with $- \log \pi_{\mathcal{\boldsymbol{\theta}}}(s_{t}|a_{t})$ to form a loss function. The model parameters $\boldsymbol{\theta}$ are then updated using gradients from this loss.

\begin{algorithm}[ht]
  \caption{Policy Gradient Algorithm for training Parametrized Quantum Circuits}\label{alg:policy_grad}
  \begin{algorithmic}[1]
    \REQUIRE Model parameters $\boldsymbol{\theta}$, Discount factor $\gamma$, Optimizer optim
    \STATE Initialize model parameters $\boldsymbol{\theta}$
    \FOR{i $\in$ [num\_episodes]}
        \STATE Generate Episode $\{s_{0}^{i}, a_{0}^{i}, r_{0}^{i}, \cdots, s_{T_{i}}^{i}, a_{T_{i}}^{i}, r_{T_{i}}^{i}\} \sim \pi_{\boldsymbol{\theta}}$
        \STATE Initialize loss $\mathcal{L}(\boldsymbol{\theta}) \leftarrow 0$
        \FOR{$t \in [T_{i}]$}
            \STATE Compute return $G_{t} = \sum_{k=0}^{T-t} \gamma^{k}r_{t+k}$
            \STATE Update loss $\mathcal{L}(\boldsymbol{\theta}) \leftarrow \mathcal{L}(\boldsymbol{\theta}) - G_{t} \log \pi_{\boldsymbol{\theta}}(a_{t}|s_{t})$
        \ENDFOR
        \STATE Update parameters $\boldsymbol{\theta} \leftarrow$ optim.update($\boldsymbol{\theta}; \nabla_{\boldsymbol{\theta}} \mathcal{L}(\boldsymbol{\theta})$)
    \ENDFOR
  \end{algorithmic}
\end{algorithm}

\subsubsection{Deep Q-Learning}

Q-Value or Action-Value function, $Q(s, a)$ in RL is the measure of expected return from a state-action pair. Similar to policy, this action value function can be parametrized by a PQC, giving the parametrized Q-Value function $Q_{\boldsymbol{\theta}}(s, a)$. The choice of action from an action-value function is often given by an $\varepsilon$-greedy selection from its Q-Value function \cite{rl_book}, i.e., we choose a random action with probability $\varepsilon$ or otherwise we choose action $a = \operatorname{argmax}_{a}Q(s, a)$. To obtain a meaningful Action-Value function from a parametrized quantum circuit, we defined $Q_{\boldsymbol{\theta}}(s, a) = \braket{O_{a}}_{s', \boldsymbol{\theta}}$ \cite{skolik2022}, where $s' = \tanh(\boldsymbol{\lambda} s)$. Algorithm \ref{alg:dq_learning} represents the Deep Q-Learning for the training of PQCs.

\begin{algorithm}[ht]
  \caption{Deep Q-Learning for training Parametrized Quantum Circuits}\label{alg:dq_learning}
  \begin{algorithmic}[1]
    \REQUIRE Model parameters $\boldsymbol{\theta}$, Discount factor $\gamma$, Optimizer optim, Loss Function $\mathcal{L}(y, \hat{y})$
    \STATE Initialize replay memory $M$ with maximum length $N$
    \STATE Initialize Action-Value function $Q_{\boldsymbol{\theta}}$ parameters $\boldsymbol{\theta}$
    \STATE Initialize target Action-Value function $\widehat{Q}_{\widehat{\boldsymbol{\theta}}}$ 
    \STATE Set $\widehat{\boldsymbol{\theta}} \leftarrow \boldsymbol{\theta}$
    \FOR{episode $\in$ [num\_episodes]}
        \STATE Sample $s_{0}$ from the environment
        \FOR{$t \in [T]$}
            \STATE Choose action $a_{t} \leftarrow \varepsilon$-Greedy Policy from $Q_{\boldsymbol{\theta}}$
            \STATE $s_{t+1}, r_{t} \leftarrow$ Next State and Reward by executing $a_{t}$
            \STATE Store the Interaction $(s_{t}, a_{t}, r_{t}, s_{t+1})$ in M.
            \STATE Sample $(s_{i}, a_{i}, r_{i}, s_{i+1})$ from M.
            \STATE Compute $\hat{y_{i}} = r_{i} + \gamma \max_{a'} \widehat{Q}_{\widehat{\boldsymbol{\theta}}}(s_{i+1}, a')$
            \STATE Compute loss $l(\boldsymbol{\theta}) = \mathcal{L}(Q_{\boldsymbol{\theta}}(s_{i}, a_{i}), \hat{y_{i}})$
            \STATE Update parameters $\boldsymbol{\theta} \leftarrow$ optim.update($\boldsymbol{\theta}; \nabla_{\boldsymbol{\theta}} l(\boldsymbol{\theta})$)
            \STATE After Every $C$ steps, Update $\widehat{\boldsymbol{\theta}} \leftarrow \boldsymbol{\theta}$
        \ENDFOR
    \ENDFOR
  \end{algorithmic}
\end{algorithm}

Apart from the description provided in Algorithm \ref{alg:dq_learning}, we used an exponential decay schedule for $\varepsilon$ starting from $\varepsilon = 1.0$ to $\varepsilon = 0.01$. This decay schedule allows the network to shift its focus from pure exploration (i.e., $\varepsilon = 1.0$) to focused exploitation, allowing room for sufficient exploration sampling for the model.

\section{Performance Evaluation}\label{sec:Experiments}

In this study, we conducted several experiments evaluating the performance of QAISim relative to the defined baseline and its classical counterparts. With careful evaluation of the task execution time, we analyze the behavior of our resource management in a quantum cloud computing environment. This data driven approach helped us making informed decisions about the model complexity and architecture to optimize quantum resource management. \revision{Furthermore, we used a two-device environment for training noisy models. Noisy models for both algorithms were a single layer PQC, trained for 150 episodes, incorporating simulated amplitude dampening and depolarization channels.}

\subsection{Experimental Setup}\label{sec:Setup}

\revision{We implemented the RL environment and the reward structure of quantum cloud computing environment for this study using OpenAI Gymnasium (Gym) \cite{gymnasium}}. All the quantum circuits were simulated using Cirq and TensforFlow Quantum. Table \ref{tab:hyprparams} shows the hyperparameters used alongwith their corresponding values for reproducing the results.

\begin{table}[ht]
    \centering
    \caption{Hyperparameters used for QAISim. lr := learning rate. Furthermore, we use update steps of 10 for the model and 30 for the target model in Deep Q-Learning.}
    \begin{tabular}{c|c|c}
        \toprule
        \textbf{Model} & \textbf{Hyperparameter} & \textbf{Value} \\
        \midrule
        \multirow{4}{*}{Policy Gradient}
        & lr$_{\boldsymbol{\phi}}$, lr$_{\mathbf{w}}$ & 0.03 \\
        & lr$_{\boldsymbol{\lambda}}$ & 0.05 \\
        & \# PQC Layers & 5 \\
        & \# Training Episodes & 1500 \\
        \midrule
        \multirow{5}{*}{Deep Q-Learning}
        & lr$_{\boldsymbol{\phi}}$, lr$_{\mathbf{w}}$ & 0.03 \\
        & lr$_{\boldsymbol{\lambda}}$ & 0.05 \\
        & \# PQC Layers & 5 \\
        & \# Training Episodes & 1500 \\
        & Decay Rate & 0.99 \\
        \bottomrule
    \end{tabular}
    \label{tab:hyprparams}
\end{table}

\subsection{Workloads/Dataset}\label{sec:Workload}

To simulate real quantum tasks in our environment, we used a subset of quantum algorithms spanning a circuit width of 2 upto 50 qubits provided by the MQT (Munich Quantum Toolkit) Benchmark Library \cite{mqt}. These tasks span a circuit depth of 2 up to 17,598 with a mean circuit layer count of 400. Furthermore, simulated 5 IBM QPUs \cite{IBMQuantum} as quantum nodes for the QRL environment. Table \ref{tab:qpu_metrics} shows the QPU metrics used for the reinforcement learning environment.

\begin{table}[ht]
    \centering
    \caption{QPU Metrics for environment quantum systems.}
    \begin{tabularx}{\linewidth}{l|X|c|c|X}
        \toprule
        \textbf{IBM QPU} & \textbf{No. of Qubits} & \textbf{EPLG} & \textbf{CLOPS} & \revision{\textbf{Mean Reward}} \\
        \midrule
        IBM Marrakesh & 156 & $3.71 \times 10^{-3}$ & 180,000 & \revision{0.0804} \\
        IBM Torino & 133 & $8.95 \times 10^{-3}$ & 200,000 & \revision{0.0894} \\
        IBM Quebec & 127 & $1.67 \times 10^{-2}$ & 32,000 & \revision{0.0143} \\
        IBM Brisbane & 127 & $1.82 \times 10^{-2}$ & 170,000 & \revision{0.0760} \\
        IBM Kolkata & 27 & $1.5 \times 10^{-2}$ & 66,000 & \revision{0.0601} \\
        \bottomrule
    \end{tabularx}
    \label{tab:qpu_metrics}
\end{table}

\begin{figure*}[ht]
   \centering
    \begin{subfigure}[c]{0.32\textwidth}
        \includegraphics[width=\linewidth]{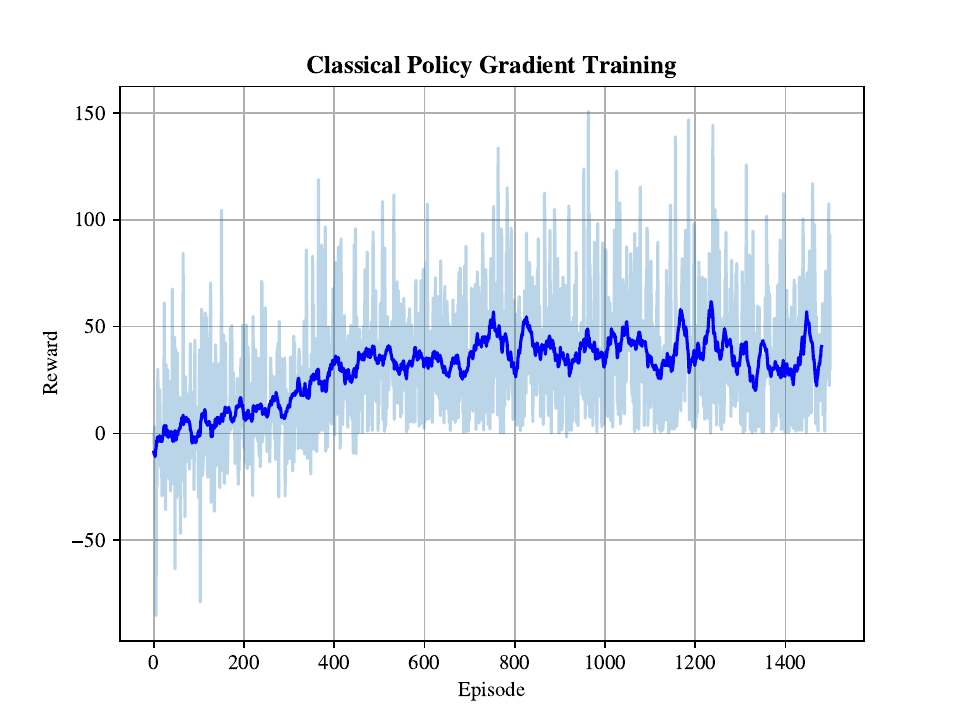}
    \end{subfigure} \hfill
    \begin{subfigure}[c]{0.32\textwidth}
        \includegraphics[width=\linewidth]{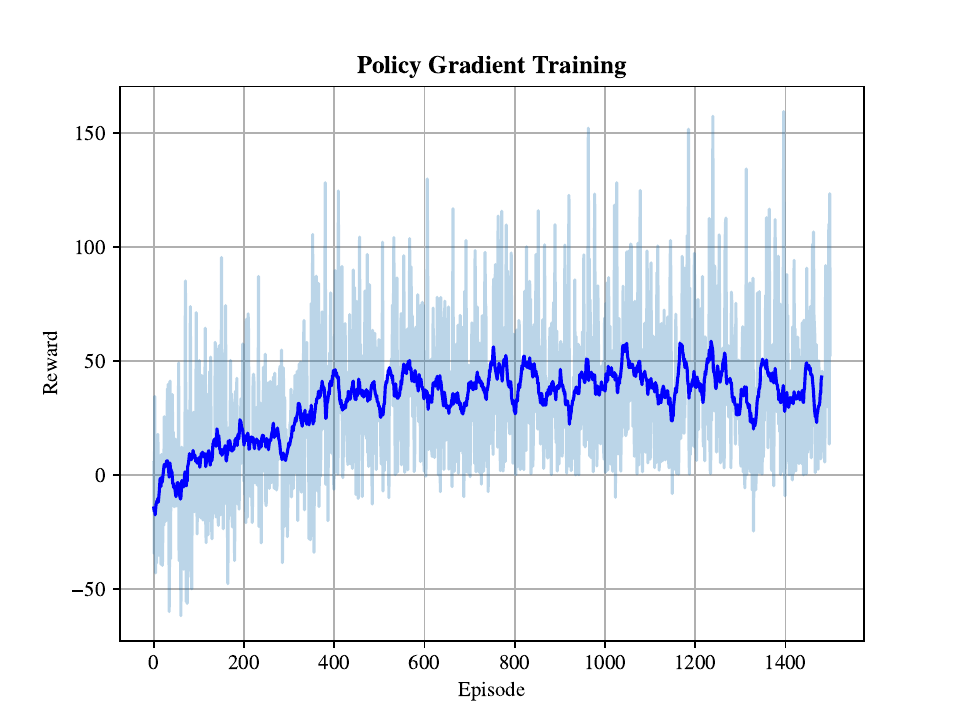}
    \end{subfigure} \hfill
    \begin{subfigure}[c]{0.32\textwidth}
        \includegraphics[width=\linewidth]{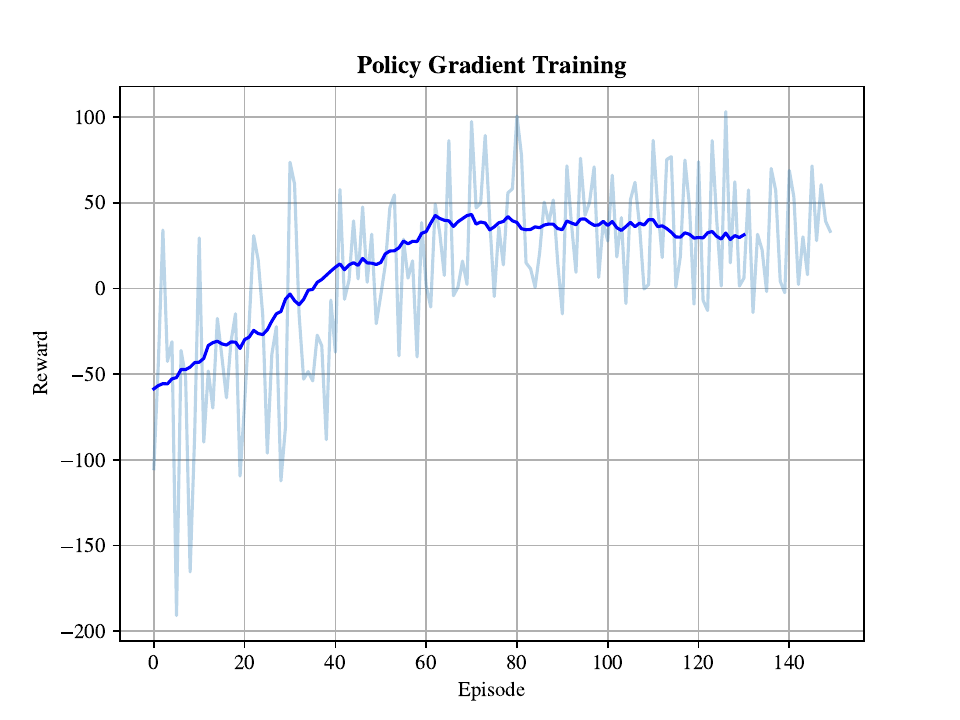}
    \end{subfigure}\\
    \begin{subfigure}[c]{0.32\textwidth}
        \includegraphics[width=\linewidth]{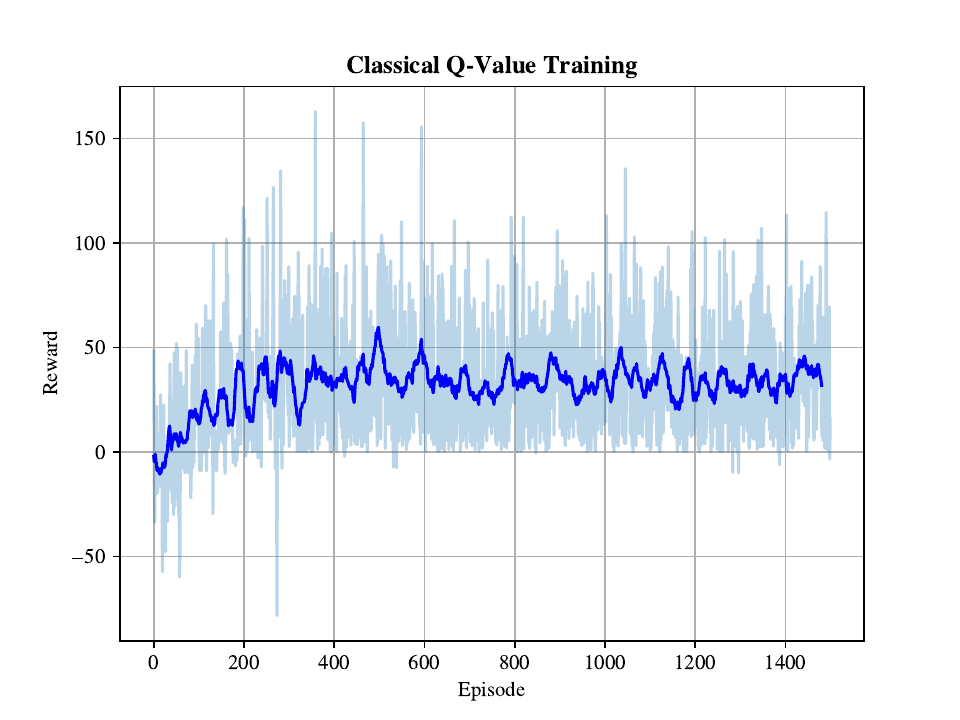}
    \end{subfigure}
    \begin{subfigure}[c]{0.32\textwidth}
        \includegraphics[width=\linewidth]{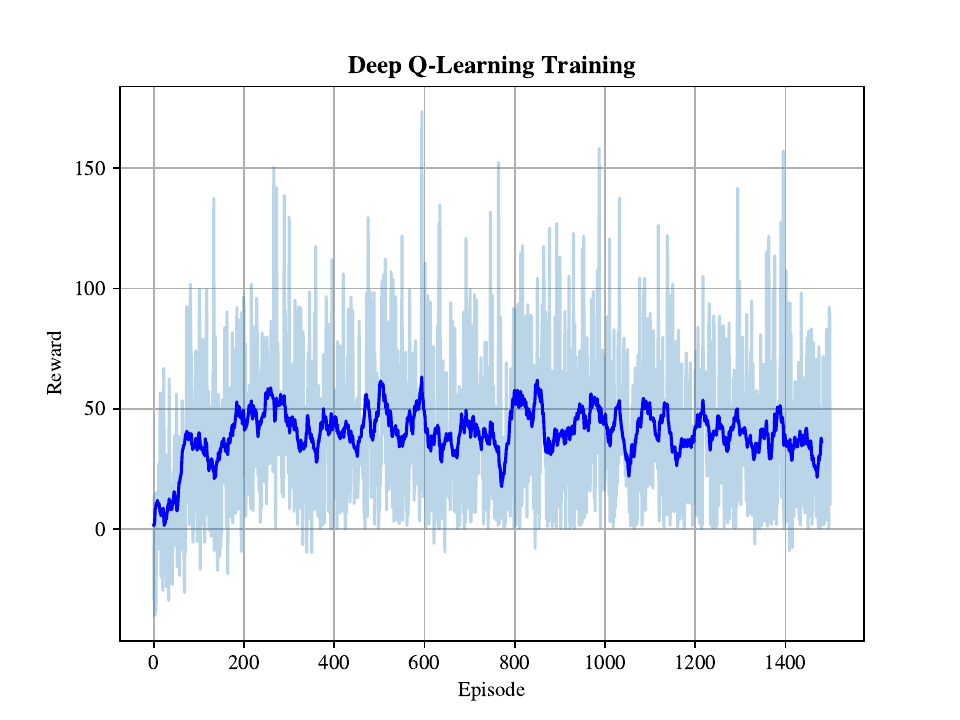}
    \end{subfigure}
    \begin{subfigure}[c]{0.32\textwidth}
        \includegraphics[width=\linewidth]{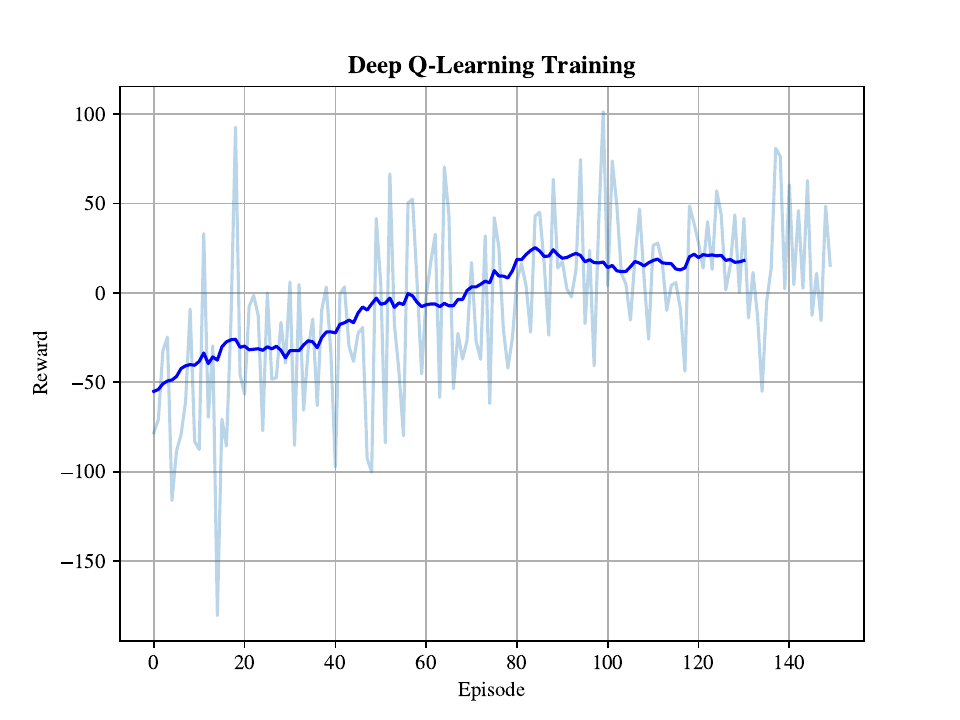}
    \end{subfigure}
    \caption{\revision{Reward-Episode distribution for training each model (classical models - left, quantum models - middle and noisy experiments - right). All the distributions are plotted by computing a moving average of 20 points for better visualization.}}
    \label{fig:model_training}
\end{figure*}

\subsection{Experimental Results}\label{sec:Results}

This section presents results for different QRL models i.e., Policy Gradient and Deep Q-Learning algongwith there classical counterparts and the baseline model using Greedy resource management algorithm.

\subsubsection{Training RL Models}\label{sec:Training}

We trained both quantum and classical reinforcement learning models using both algorithms i.e., policy gradient and deep Q-learning. Section \ref{sec:Setup} discusses the architectural design of parametrized quantum circuits used for QAI models. We used a 3 layered MLP for the classical DRL with hidden layers of 64 dimension. We trained both models (quantum and classical) on same environments and training of both the models converged with similar rewards. Figure \ref{fig:model_training} shows the distribution of rewards with each episode for all the models. For both quantum and classical models, we used Adam optimizer with learning rate of $1 \times 10^{-3}$ for classical models.

\subsubsection{Evaluation}

\begin{figure}[ht]
    \centering
    \begin{subfigure}[c]{0.48\textwidth}
        \includegraphics[width=\linewidth]{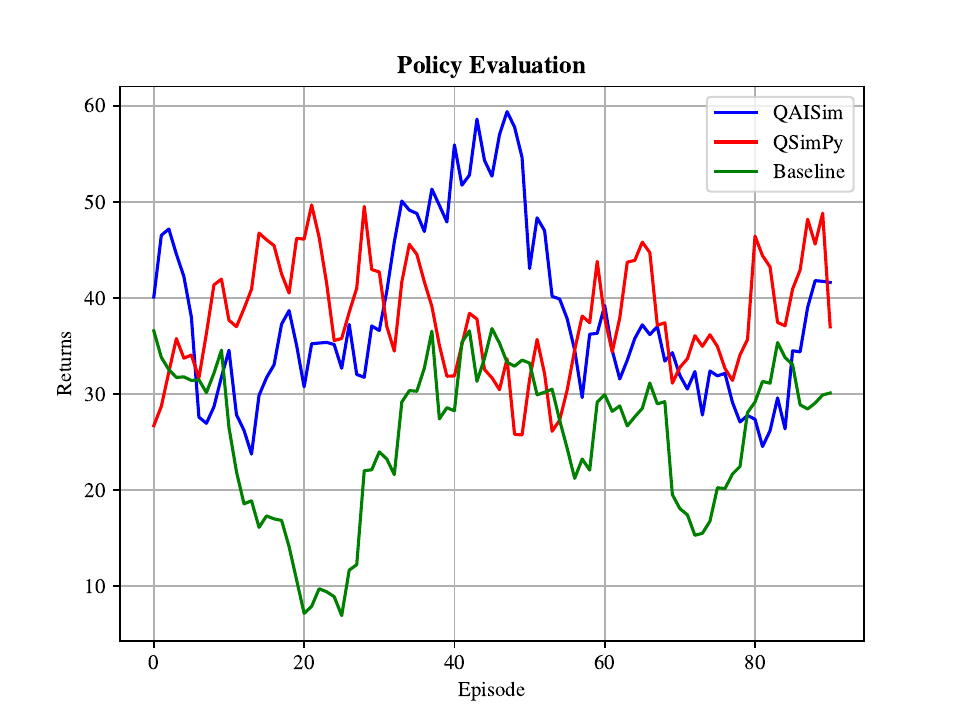}
    \end{subfigure} \hfill
    \begin{subfigure}[c]{0.48\textwidth}
        \includegraphics[width=\linewidth]{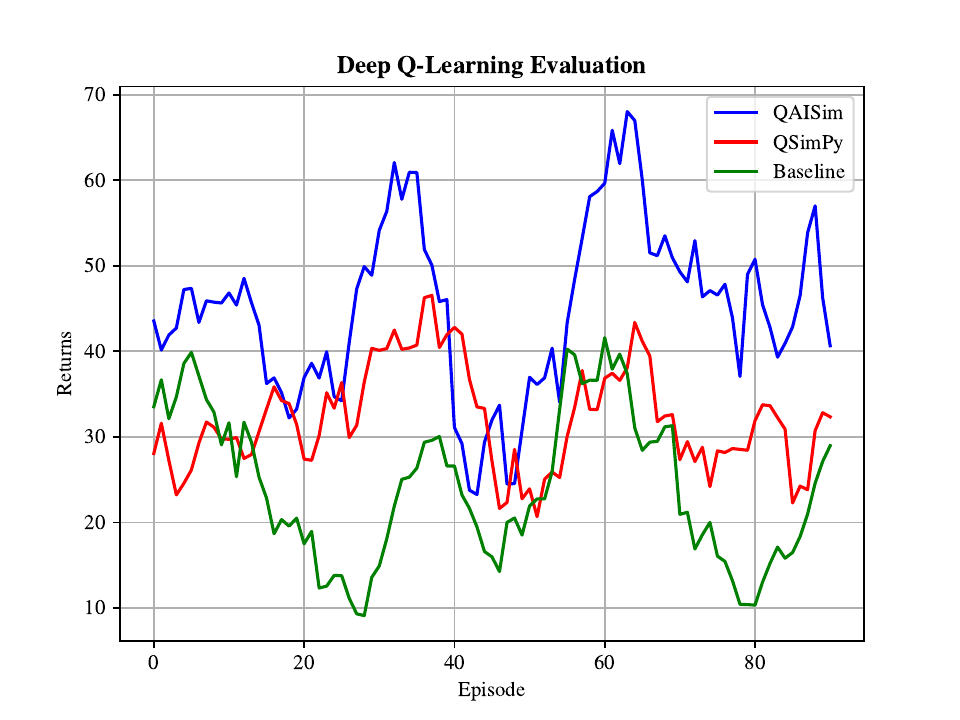}
    \end{subfigure}
    \caption{\revision{Cumulative per-episode returns from the environment for all models and the baseline greedy model. Both evaluation plots are plotted after taking a moving average of 10 points for better visualization.}}
    \label{fig:eval_returns}
\end{figure}

\begin{figure}[ht]
    \centering
    \begin{subfigure}[c]{0.48\textwidth}
        \includegraphics[width=\linewidth]{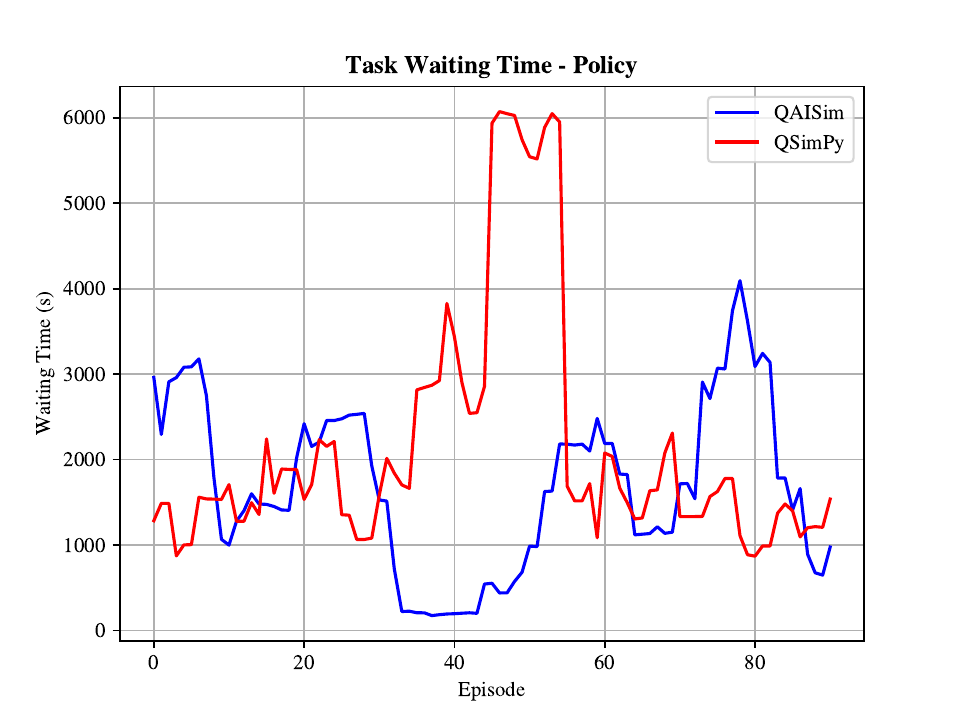}
    \end{subfigure} \hfill
    \begin{subfigure}[c]{0.48\textwidth}
        \includegraphics[width=\linewidth]{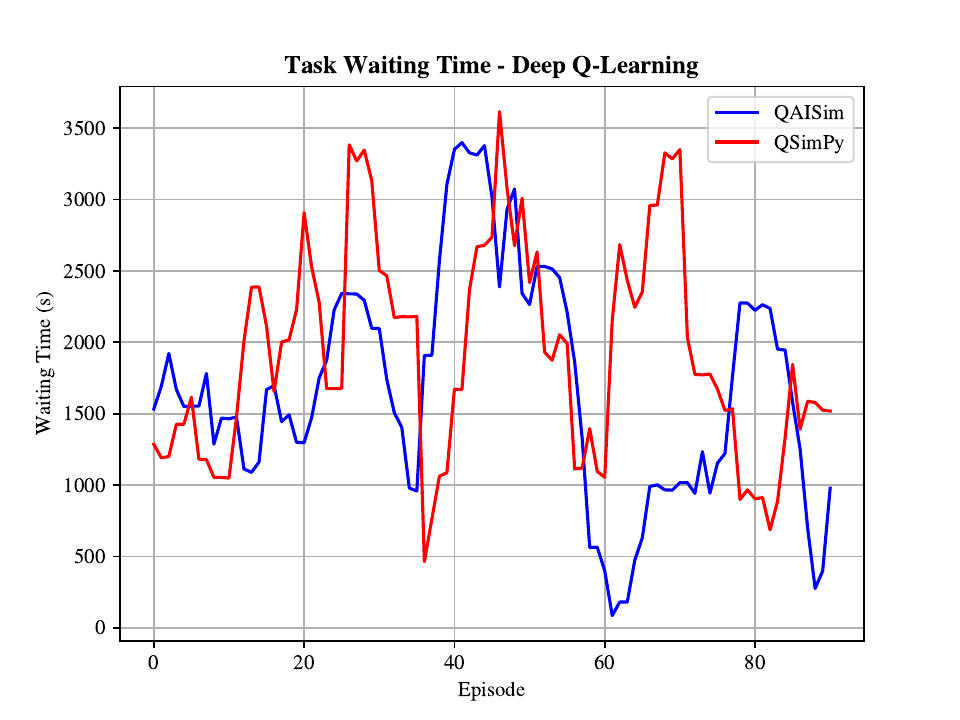}
    \end{subfigure}
    \caption{\revision{Cumulative per-episode waiting time from the environment for all models. Both evaluation plots are plotted after taking a moving average of 10 points for better visualization.}}
    \label{fig:eval_wt}
\end{figure}

To evaluate our reinforcement learning models and the baseline, we used cumulative returns from the QRL environment and the cumulative task waiting time for each episode. Figure \ref{fig:eval_returns} shows the evaluated cumulative per-episode returns for all models, and Figure  \ref{fig:eval_wt} shows the cumulative per-episode waiting time for all models. The observed comparable performance of quantum RL models with classical ones shows the capabilities of the QAI models to compete with classical AI models. Table \ref{tab:eval} additionally shows that the quantum models’ average per‑episode returns fall within approximately 10\% of their classical counterparts, indicating comparable resource‑management performance.

\begin{table}[ht]
    \centering
    \caption{Average cumulative per-episode returns for all models on randomly sampled tasks.}
    \begin{tabular}{c|c|c}
        \toprule
        \textbf{Training Algorithm} & \textbf{Model} & \textbf{Average Returns} \\
        \midrule
        \multirow{3}{*}{Policy Gradient}
        & Baseline \cite{baseline} & 29.3311 \\
        & QSimPy \cite{qsimpy} & 40.0606 \\
        & \textbf{QAISim} & \textbf{44.0417} \\
        \midrule
        \multirow{3}{*}{Deep Q-Learning}
        & Baseline \cite{baseline} & 25.5270 \\
        & QSimPy \cite{qsimpy} & 45.3555 \\
        & \textbf{QAISim} & \textbf{40.0432} \\
        \bottomrule
    \end{tabular}
    \label{tab:eval}
\end{table}

Noticeably, the quantum models of QAISim demonstrate better resource management capabilities in contrast to the baseline greedy algorithm. This positive result can be credited to the adoption of parametrized quantum circuits, which leverage reinforcement learning and quantum computing methodologies. As a result, QAISim has effectively presented the use of quantum machine learning within the realm of cloud computing. Furthermore, QAISim demonstrates advantage in number of trainable parameters compared to classical reinforcement learning as the number of trainable parameters increase linearly with each layer in a PQC. \revision{For reference, the 5-layer PQC underlying both algorithms consist of 181 trainable parameters. Whereas, the 3-layer MLP underlying both classical models consists of 9,221 trainable parameters.}

\section{Conclusions and Future Scope}\label{sec:Conclusion}
The rapid advancement in quantum computing technologies has rendered traditional quantum circuit simulations insufficient for the needs of the quantum computing community. Quantum cloud computing provides access to quantum resources for testing and executing quantum tasks via a cloud-based interface, enabling scalable, remote collaboration and development in real-time. However, these cloud-based solutions have limited resources, necessitating a need for efficient resource management to fulfil the growing computational demand of modern IoT applications. In this paper, we have proposed QAISim, a novel framework that leverages QAI models to address efficient resource management in quantum cloud computing. By integrating advanced RL algorithms such as Policy Gradient and Deep Q-Learning with quantum computing, we have successfully demonstrated the capability of QAISim to perform efficient resource management on par with the classical models. In summary, our work showcases the potential of QAISim as a powerful tool for quantum cloud resource management by leveraging advanced QAI models.

\subsection{Promising Future Directions}
While our work primarily focuses on efficient resource management to reduce overall task execution time by proposing a basic version of QAISim, it provides a strong foundation for several promising avenues of research and development.

\revision{\subsubsection{Multi-Objective Optimization}
Moving beyond task execution time, researchers can explore multi-objective optimization using QAISim, leveraging intelligently designed reward functions to balance multiple goals like execution time, energy efficiency, security, among other factors to provide an optimized solution \cite{multi_objective}. This extension of QAISim framework will be essential for real quantum cloud systems, where diverse constraints needs to be satisfied while maintaining optimal performance.}

\subsubsection{Experimental Evaluation on Quantum Hardware}
We currently simulate all QAI models in TensorFlow, which integrates the quantum circuit computations into the automatic differentiation computation graph. Going forward, researchers can extend the QAISim framework by training and deploying QAI models on real quantum hardware using quantum-based gradient computation \cite{q_grad}. This way, researchers could evaluate true quantum-accelerated learning and inference in practice.

\subsubsection{Artificial Intelligence of Things (AIoT)}
Modern AIoT applications, such as healthcare, agriculture, and transportation, can integrate QAISim with edge computing to offer low-latency benefits to end devices \cite{hatay2025transforming}. Incorporating these aspects into the development and enhancement of QAISim will ensure its relevance and effectiveness in addressing the resource management challenge in the growing field of quantum cloud computing.

\section*{Software Availability}
We released QAISim, available for free as open source. All code, datasets, and result reproducibility scripts are publicly available and can be accessed from GitHub:
\url{https://github.com/Irwindeep/QAISim} 

\section*{Acknowledgments}
The research investigation was carried out during the internship undertaken by Irwindeep Singh at School of Electronic Engineering and Computer Science, Queen Mary University of London, UK. 

\section*{Declarations}
\begin{itemize}
\item Funding \\
NA
\item Conflict of interest/Competing interests \\
On behalf of all authors, the corresponding author states that there is no conflict of interest.
\item Ethics approval 
\\
Not Available 
\item Consent to participate
\\
Not Available 
\item Consent for publication
\\
Not Available 
\item Availability of data and materials
\\
Not Available 
\item Code availability 
\\
The code is available at \url{https://github.com/Irwindeep/QAISim} 
\item Authors' contributions\\
\textbf{ Irwindeep Singh } (Conceptualization: Lead; Data curation: Lead; Formal analysis: Lead; Funding acquisition: Lead; Investigation: Lead; Methodology: Lead; Software: Lead; Validation: Lead; Writing – original draft: Lead)
\textbf{ Sukhpal Singh Gill } (Conceptualization: Lead; Investigation: Lead; Methodology: Lead; Validation: Lead; Writing – original draft: Lead; Supervision: Lead)
\textbf{ Jinzhao Sun} (Conceptualization: Lead; Formal analysis: Lead; Writing – original draft: Lead; Writing – review \& editing: Supporting) 
\textbf{ Jan Mol } (Conceptualization: Lead; Formal analysis: Lead; Writing – original draft: Lead; Writing – review \& editing: Supporting) 
\end{itemize}

\small
\bibliographystyle{IEEEtran}
\bibliography{ref}

\section*{Biographies}
\vspace{-0.6in}
\begin{IEEEbiography}[{\includegraphics[width=1in,height=1in,clip]{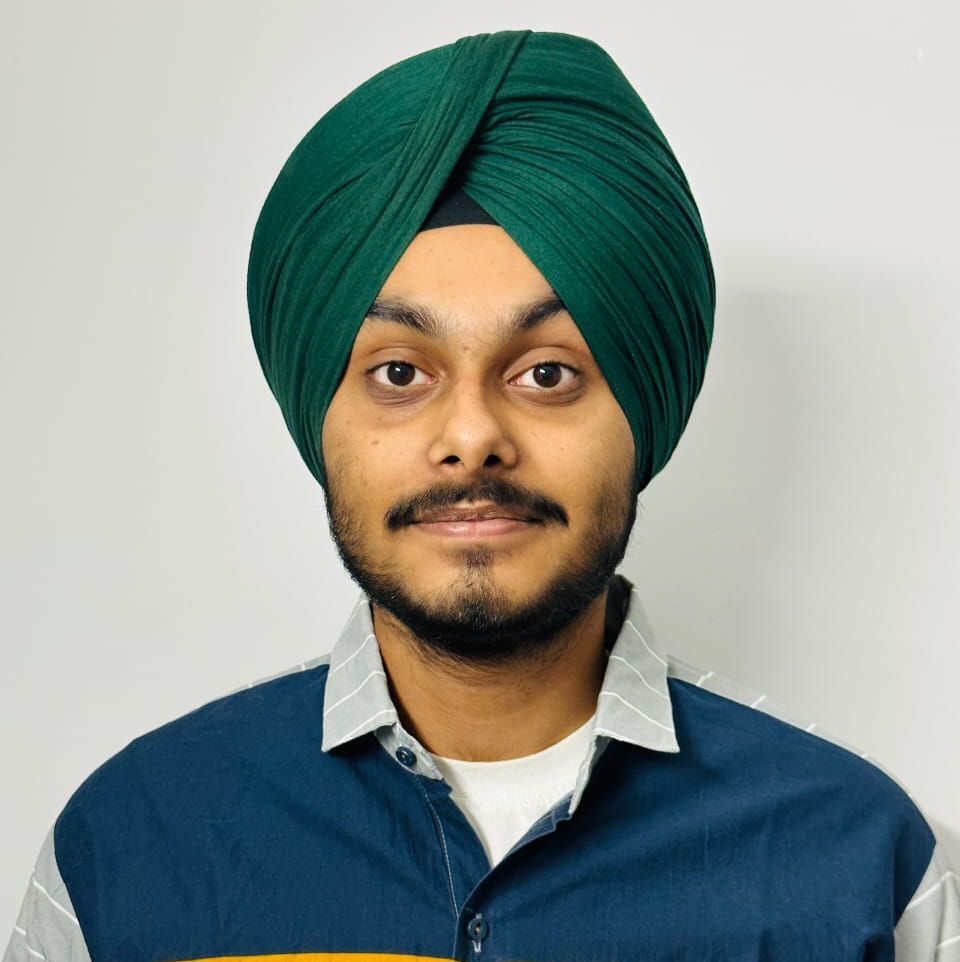}}]{Irwindeep Singh} is an undergraduate student at Department of Computer Science and Engineering at Indian Institute of Technology, Jodhpur, India. He is working as a research intern at Next Generation InteLLigent Systems and Networks (GiLLNet) Lab, School of Electronic Engineering and Computer Science, Queen Mary University of London. His research interests include Machine Learning, Deep Learning, Internet of Things (IoT) and Quantum Computing. 
\end{IEEEbiography}
\vspace{-0.6in}
\begin{IEEEbiography}[{\includegraphics[width=1in,height=1.15in,clip]{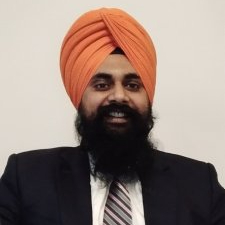}}]{Sukhpal Singh Gill}  (SFHEA) is an Assistant Professor of Cloud Computing at the School of Electronic Engineering and Computer Science, Queen Mary University of London, UK. Dr. Gill is serving as an Editor-in-Chief for IGI Global IJAEC and Area Editor for Springer Cluster Computing Journal, also serving as an Associate Editor in IEEE IoT, Nature Scientific Reports, Wiley SPE, Elsevier IoT, Wiley ETT, Elsevier Telematics and Informatics Reports and IET Networks Journals. He has co-authored 200+ peer-reviewed papers (with Citations 15700+ and H-index 62) and has published in prominent international journals and conferences such as IEEE TPDS, IEEE TFS, IEEE TGCN, IEEE TCC, IEEE TSC, IEEE TSUSC, IEEE COMST, IEEE TCE, ACM TOIT, ACM CSUR, ACM TAAS, IEEE TII, IEEE TNSM, IEEE IoT Journal, Elsevier JSS/FGCS, IEEE/ACM UCC and IEEE CCGRID. He has received several awards, including the Queen Mary University Education Excellence Award 2023, Outstanding Reviewer Award from IEEE IT Professional Magazine 2024, Elsevier Internet of Things Editor’s Choice Award 2024, Elsevier Best Paper Award 2023, Distinguished Reviewer Award from SPE (Wiley), Best Paper Award AusPDC at ACSW 2021. He has edited and authored research various books for Elsevier, Springer and CRC Press. His research interests include Cloud Computing, Edge Computing, IoT and Energy Efficiency. For further information, please visit: \url{http://www.ssgill.me}.
\end{IEEEbiography}
\vspace{-0.6in}
\begin{IEEEbiography}[{\includegraphics[width=1in,height=1.1in,clip]{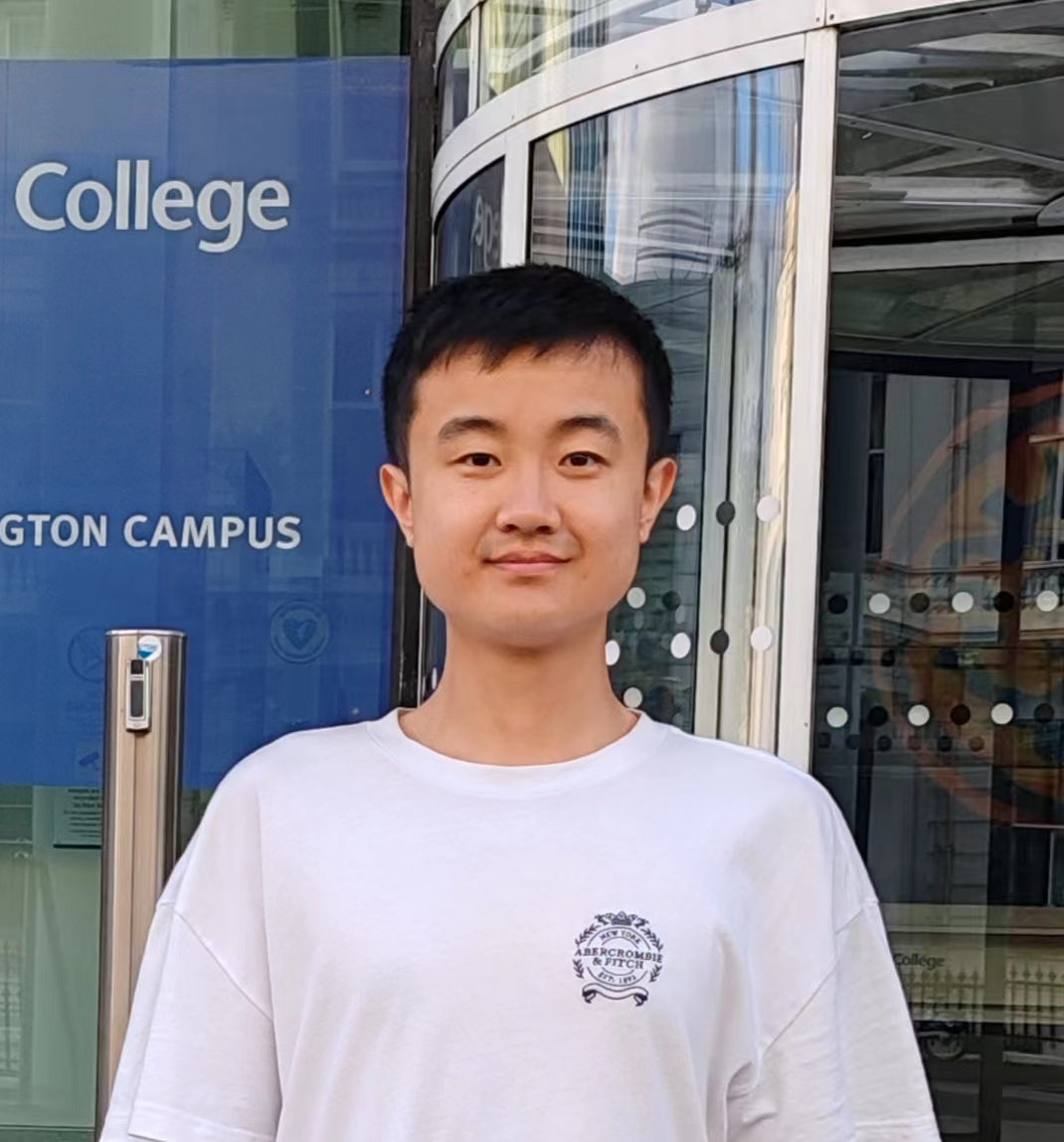}}]{Jinzhao Sun} is an Assistant Professor at Queen Mary University of London. Prior to this, he received the Schmidt AI in Science fellowship and Awards for Excellence at the University of Oxford, where he holds a research fellowship and serves as a College Advisor at Reuben College. He has a track record of publishing in prestigious journals such as Nature Physics, Nature Communications and Physical Review Letters. His current research interest focuses on quantum algorithms, quantum simulation and many-body quantum dynamics.
\end{IEEEbiography}
\vspace{-0.6in}
\begin{IEEEbiography}[{\includegraphics[width=1in,height=1.33in,clip]{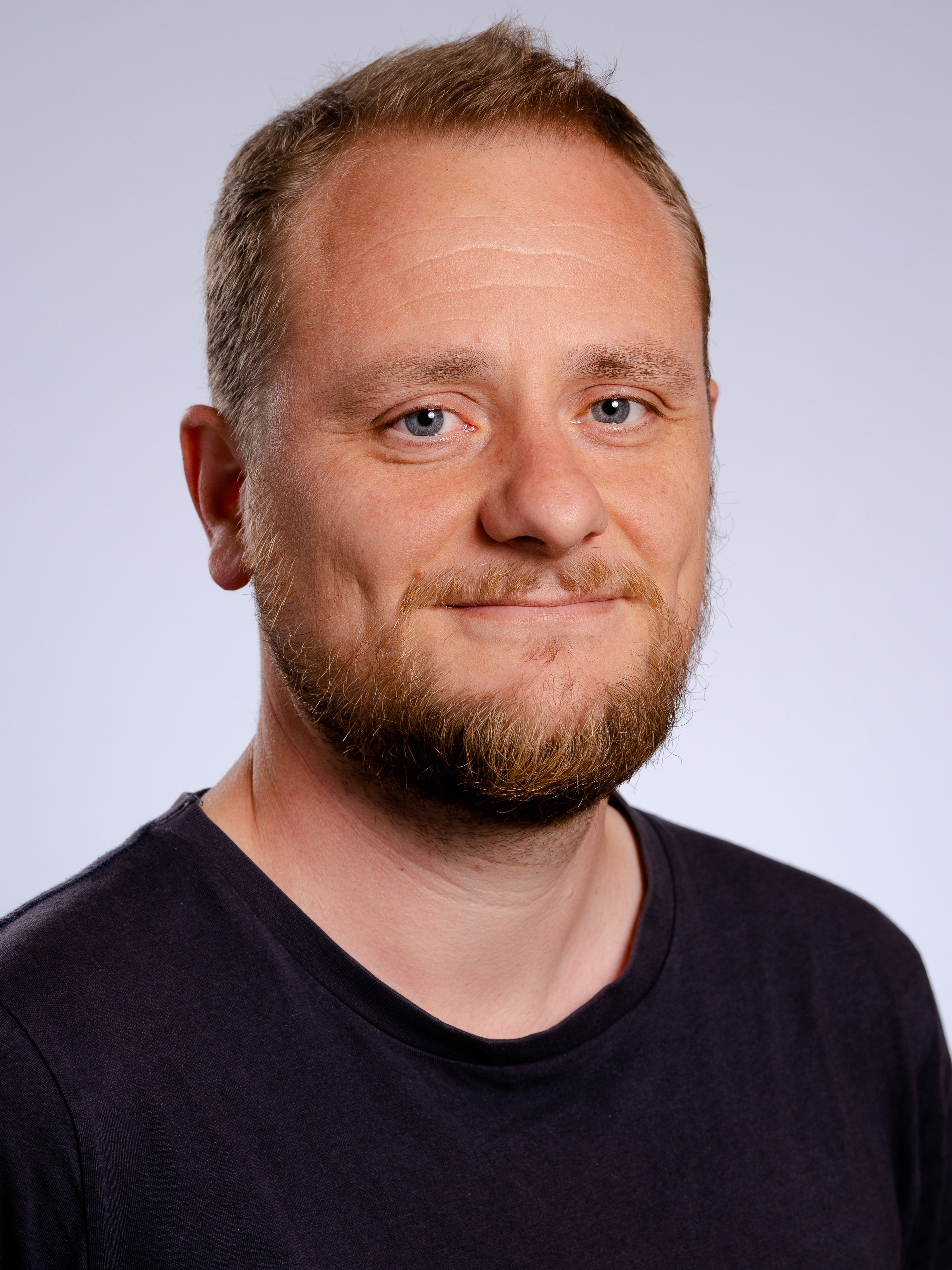}}]{Jan Mol} has a strong track record of publishing in high-impact journals such as ACS Nano and Nature Materials. His ability to lead significant independent research has been recognised by the Royal Society, the Royal Academy of Engineering, and UKRI, who have each awarded him with a research fellowship. Jan has also demonstrated his ability to collaborate and manage strongly multidisciplinary teams, specifically as co-investigator on the EPSRC QuEEN Programme Grant. His latest research fellowship – the UKRI Future Leaders Fellowship – has enabled Jan to build on his early-career successes, which include fruitful collaborations with industrial partners and the establishment of international network supported by the Global Challenges Research Fund, and accelerate his research; he is bringing together an excellent  team of postdocs and PhD students with complementary skills in physics, chemistry, and materials science to tackle challenges in energy, health, and information processing. To achieve his goals, Jan collaborates with academic and industrial partners from across the UK and world-wide (e.g. USA, Singapore, Greece, and South Africa). Finally, Jan has taken a leading role in engaging with wider audiences by organising workshops and symposia, and by developing outreach materials and activities related to atomic- and molecular-scale devices.
\end{IEEEbiography}

\end{document}